\documentclass[preprint,showpacs,showkeys,aps]{revtex4}
\usepackage[english]{babel}

\usepackage[dvips]{graphicx}
\usepackage{amssymb}

\begin{document}

\title{
Heat Conduction, and the Lack Thereof, in Time-Reversible
Dynamical Systems:
Generalized Nos\'e-Hoover Oscillators
with a Temperature Gradient
}

\author{Julien Clinton Sprott}

\affiliation{
Department of Physics \\
University of Wisconsin \\
Madison, Wisconsin 53706 \\
}

\author{William Graham Hoover and Carol Griswold Hoover}

\affiliation{
Ruby Valley Research Institute                  \\
Highway Contract 60, Box 601                    \\
Ruby Valley, Nevada 89833                       \\
}

\date{\today}

\vspace{0.1cm}

\begin{abstract}

We use nonequilibrium molecular dynamics to analyze and
illustrate the qualitative differences between
the one-thermostat and two-thermostat versions of equilibrium
and nonequilibrium (heat-conducting) harmonic oscillators.
Conservative nonconducting regions can coexist with
dissipative heat conducting regions in phase
space with exactly the same imposed temperature field.

\end{abstract}

\maketitle

\section{Hamiltonian Background}
In 1984, Shuichi Nos\'e modified Hamiltonian mechanics to make it
consistent with Gibbs' canonical isothermal distribution rather than the usual
microcanonical isoenergetic one\cite{b1,b2}.  This achievement was a crucial step toward
reconciling time-reversible microscopic mechanics with macroscopic (irreversible)
thermodynamics.  The time-reversible nature of the work we describe here owes its
origin to elaborations of Nos\'e's Hamiltonian research\cite{b3,b4,b5,b6} and its
relationship to the Second Law of Thermodynamics\cite{b7}. Sprott discovered the
special case detailed and embellished upon here through an independent, and quite different,
approach\cite{b8,b9}.  In their 1986 paper Politi, Oppo, and Badii investigated 
laser systems described by three-equation models similar to the thermostated oscillator systems
treated here\cite{b10}.

Shortly after Nos\'e's seminal (equilibrium) papers, it was discovered that his
time-reversible equations {\it can} lead directly to models for irreversible behavior,
providing a computer-age rejoinder to Loschmidt's Paradox\cite{b7}.  In the language
of nonlinear dynamics, the {\it time-reversible} equations (in the sense that a
reversed movie satisfies the same equations) lead to a {\it dissipative} contracting
phase-space flow {\it from} a multifractal repellor {\it to} the repellor's mirror image,
$\{ \ +p \longleftrightarrow - p \ \}$, a strange attractor.

A variety of problems were studied to help ``understand'' the consequences of
Nos\'e's work for irreversible flows.  Some of these used periodic boundaries\cite{b7}, while others
used separate ``cold'' and ``hot'' reservoir regions\cite{b6}.  Typically these systems had
relatively complex behavior and are thus not a subject of current research.  To illustrate Nos\'e's
work, we consider here the ``simplest'' interesting application, a one-dimensional harmonic
oscillator\cite{b3,b4,b5,b6,b8,b9,b11,b12}.  The {\it nonequilibrium} dynamics of this oscillator is
relatively complicated compared with its equilibrium counterpart\cite{b3,b4}.  The modified
oscillator Hamiltonian includes Nos\'e's new ``time-scaling'' variable $s$ along with its conjugate
momentum $p_s$ and a fixed thermodynamic temperature $T$:
$$
2{\cal H} = (1/m)(p/s)^2 + \kappa q^2 + (1/M)p_s^2 + kT\ln(s^2) \ .
$$
To simplify notation, we choose $(\ m,\kappa,M,k \ )$ all equal to unity in what
follows, so that the modified equations of motion are:
$$
\{ \ \dot q = (p/s^2) \ ; \ \dot p = -q \ ; \
\dot s = p_s \ ; \ \dot p_s = (p^2/s^3) - (T/s) \ \} \ .
$$
Nos\'e introduced an unusual trick, multiplying all the time derivatives
by $s$ , which he called ``scaling the time''\cite{b1,b2,b3,b4}.  He then replaced $p/s$ with $p$
to obtain:
$$
\{ \ \dot q = p \ ; \ \dot p = -q - p_sp \ ; \ \dot p_s = p^2 - T \ ; \ \dot s = sp_s \ \} \ .
$$
Because $s$ plays no role in the evolution of the other variables, Hoover suggested
omitting it and replacing the residual ``momentum''  $p_s$ by a
``friction coefficient'' $\zeta$ \cite{b3}.

Hoover also provided a simpler derivation of the ``Nos\'e-Hoover'' motion equations and
pointed out that it is easy to verify that a generalized version of the canonical
phase-space probability density is a stationary solution of the phase-space continuity
equation coupled with the Nos\'e-Hoover equations of motion:
$$
f(q,p,\zeta) \propto e^{-q^2/2T}e^{-p^2/2T}e^{-\zeta^2/2T}
 \stackrel
{(\dot q,\dot p,\dot \zeta)}{\longleftrightarrow}
(\partial f/\partial t) \equiv 0 \ .
$$
This observation provides the simplest derivation of the Nos\'e-Hoover motion equations:
assume that the distribution has the desired form and use that assumption to find
consistent equations of motion by applying the continuity equation.  Bauer, Bulgac,
and Kusnezov carried out a systematic exploration of thermostated equations of motion
based on this probability-density approach\cite{b13}.  Among their many results, two stand
out: [1] {\it cubic} frictional terms (such as $-\zeta p^3 \ {\rm or} \ -\zeta^3p$
as discussed further here) facilitate ergodicity; [2] with {\it three}
fully time-reversible control variables (``demons'' in the BBK terminology), even
Brownian motion can be simulated with time-reversible mechanics.

A dozen years later, Dettmann discovered that these same motion equations follow
from a slightly-different Hamiltonian\cite{b14,b15}, provided this alternate Hamiltonian is set
equal to zero:
$$
\{ \ \dot q = p \ ; \ \dot p = -q - \zeta p \ ; \ \dot \zeta = p^2 - T \ \} \longleftarrow
{\cal H}_{\rm Dettmann} \equiv s{\cal H}_{\rm Nos\acute{e}} \equiv 0 \ [ \ ! \ ] \ .
$$

\section{Nonequilibrium Applications of Nos\'e's Ideas}

The Second Law of Thermodynamics affirms that nonequilibrium systems generate entropy
in themselves or in their surroundings.  Nos\'e-Hoover mechanics identifies this Gibbsian
statistical-mechanical entropy with the heat extracted by the friction coefficient(s)
$\{ \ \zeta(T) \ \}$, divided by the corresponding temperature(s) $T$.  Identifying the microscopic
$\langle \ p^2 \ \rangle $ with the macroscopic temperature gives the usual thermodynamic
relation linking entropy to heat flow.  Furthermore, it is established that the friction coefficient
$\zeta $ is the rate of entropy extraction from the system.  If the system is in a steady
nonequilibrium state, then this extraction rate equals the
rate of entropy production within the system:
$$
\langle \ \dot S = \dot Q/T = \zeta (p^2/T) = \zeta \ \rangle_{\rm Steady \ State} \ .
$$
There is a voluminous literature detailing applications of Nos\'e-Hoover mechanics to heat
extraction from systems undergoing nonequilibrium flows of mass, momentum, and energy.

It is less well-known that Hamiltonian systems are unable to play this r\^ole of linking
the microscopic and macroscopic descriptions of dissipation.  In typical situations,
where conservative Lagrangian/Hamiltonian mechanics is used to constrain the temperature of
selected degrees of freedom, the conservative nature of the mechanics prevents heat transfer,
so that one can generate systems with huge temperature gradients (imposed by
hot and cold constraints on selected degrees of freedom described by Hamiltonian mechanics),
which nevertheless transmit no heat\cite{b16}.

The goal of this paper is to study the detailed dynamics of simple oscillator
systems in a nonequilibrium thermal environment to elucidate
the link established by Nos\'e between microscopic mechanics/dynamics and macroscopic
irreversible thermodynamics.  In the following two sections we give detailed numerical
studies of heat transfer, or its lack, in these nonequilibrium oscillator systems.
Afterward, we point out additional problems not addressed here but worthy of study.
Finally, we summarize the conclusions.

\section{Heat Transfer, or {\it not}, with a Harmonic Oscillator}

From the standpoint of dynamics, a bare-bones parsimonious equilibrium system (free
of time-averaged dissipation) is a thermostated harmonic oscillator,
$$
\{ \ \dot q = p \ ; \ \dot p = -q - \zeta p \ ; \ \dot \zeta = p^2 - T \ \} \ .
$$
Sprott found these same equations with an automated computer search for simple
dynamical equations displaying chaos\cite{b8}, and it was the simplest time-reversible system found.
When the initial conditions are specified, the
equilibrium oscillator maps out a portion of the $(\ q,p,\zeta \ )$ phase
space consistent with these initial conditions.  When the temperature $T$ is constant, a
wide variety of stationary phase-volume conserving solutions are obtained by varying the
initial conditions.  The set of initial conditions $(\ q,p,\zeta \ ) = (\ 0,5,0 \ )$,
generates a chaotic sea, a contiguous Lyapunov unstable region perforated by an
infinity of quasi-periodic toroidal orbits.  By contrast, the initial conditions
$(\ q,p,\zeta \ ) = (\ 1,0,0 \ )$ generate a simple torus.  If the response
rate of the thermostat $\dot \zeta$ is scaled by a relaxation time $\tau$:
$$
\dot \zeta = [ \ p^2 - T \ ]/\tau^2 \ ,
$$
an infinite variety of different equilibrium solutions can be obtained, with the
complexity of the structures increasing as $\tau$ approaches zero\cite{b4,b6}. Figure
1 shows three equilibrium cross sections from the chaotic sea as well as two
tori and the projection of one of them onto the $(q,p)$ plane.  These solutions are not
dissipative, and they obey a time-averaged version of the equilibrium Liouville's Theorem,
$\langle \ (d\ln f/dt) \ \rangle \equiv 0$.

\begin{figure}[!ht]
\centering
\includegraphics[scale=0.5]{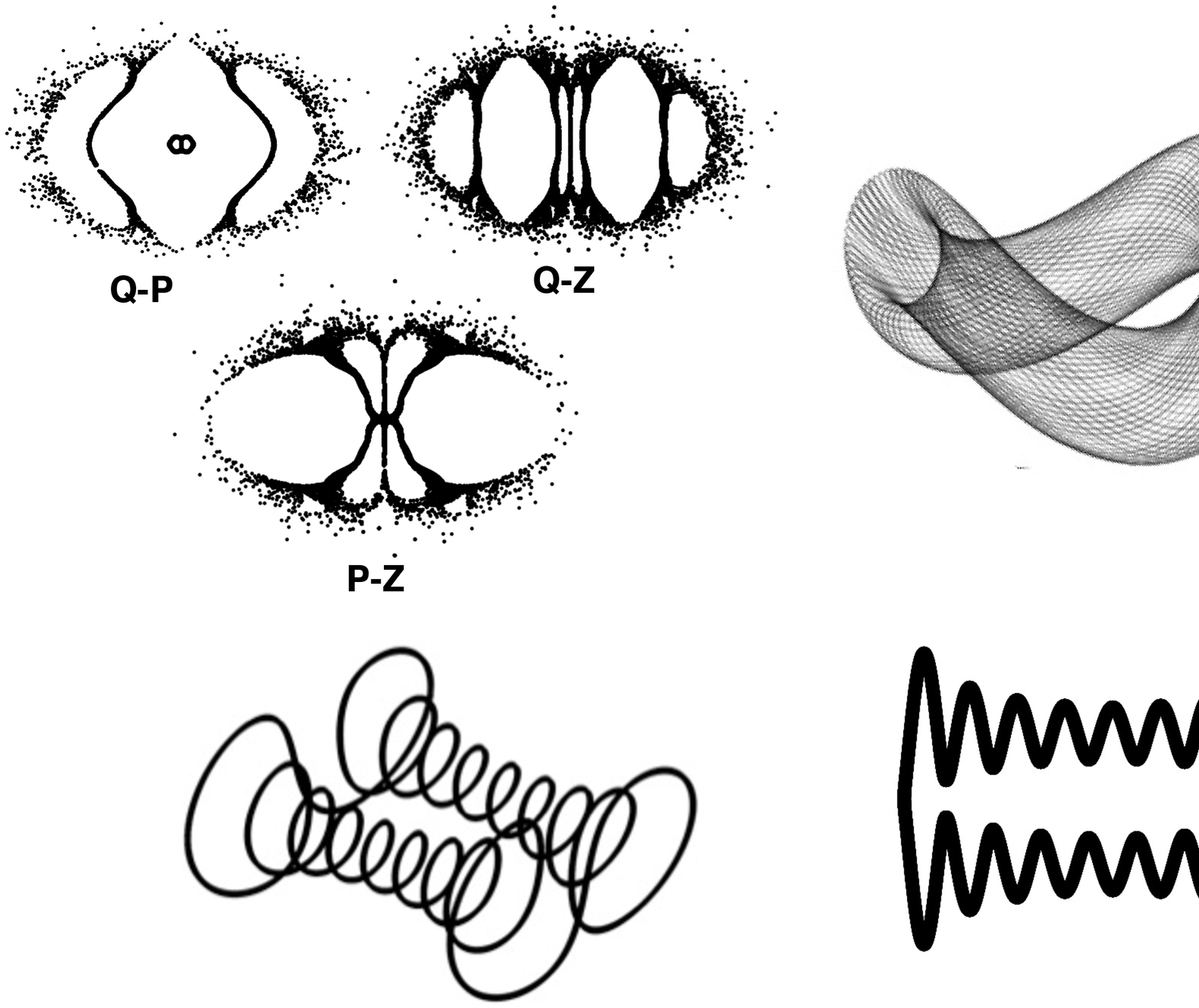}
\caption{
Chaotic and torus solutions of the equilibrium Nos\'e-Hoover equations: \\
$\{ \ \dot q = p \ ; \ \dot p = -q -\zeta p \ ; \ \dot \zeta \propto p^2 -1 \ \} \ $.
See Reference 4 for more details.
}
\end{figure}

Consider next the nonequilibrium situation where the imposed temperature field is a
function of the oscillator coordinate $q$.  We use a simple interpolation between
$T(\mp \infty) = 1 \mp \epsilon$ :
$$
T(q) = 1 + \epsilon \tanh(q) \ .
$$
Here $\epsilon$ is the maximum value of the temperature gradient, $(dT/dq)$, which occurs at $q=0$\emph{}.
One would expect the dissipation to cause the oscillator to
transfer heat in the direction counter to the gradient.  In a {\it dissipative} solution,
there is a persistent loss of phase volume, $\langle \ \dot \otimes \ \rangle < 0$. This
loss, along with chaos, is the identifying characteristic of a strange attractor with a
fractional information dimension.

Accurate solutions of the temperature-dependent motion equations testing this idea of
irresistible dissipation (and often verifying it) can be obtained by numerical
integration. For sufficiently-large gradients the oscillator typically has a limit cycle in which
heat is transferred in the negative $q$ direction.  For smaller gradients, the details
become messy, with the same sort of intricate Poincar\'e structures associated with
Hamiltonian chaos.  At $\epsilon = 0.45$ most trajectories collapse to a limit
cycle, while at $\epsilon = 0.40$ there is a stable torus solution.  In addition
to these relatively robust structures, holes in the strange-attractor cross sections
for $0 < \epsilon < 0.40$ indicate the locations of the (infinitely-many)
quasi-periodic solutions which thread through the chaotic region.  Careful work locates
the transition linking the chaotic and limit-cycle solutions near $\epsilon = 0.4053_6$ .

The two-dimensional sections of the three-dimensional flow are dazzling in their complexity.
Their analysis yields a hybrid surprise: parts of the nonequilibrium phase space are, as
expected, dissipative, with $\langle \ \zeta \ \rangle > 0$ ; but {\it other} parts,
invariant tori, are {\it conservative}, with $\langle \ \zeta \ \rangle = 0$.  We found it
puzzling that these tori are not dissipative.

A helpful referee pointed out to us that
a very similar coexistence of conservative and dissipative solutions was found in the
laser models investigated in Reference 10.
A little reflection suggests that a torus
mapping into itself must obey Liouville's Theorem and so cannot shrink.
 If dissipation were possible it would
necessarily cause the tori to shrink, until they reached their limit cycles.  On the
other hand, by applying a sufficiently large temperature gradient ($\epsilon = 0.45$ is
large enough), {\it all} the tori can be made to disappear. This remarkable feature,
where conservative states coexist with dissipative states is possible
because the damping is {\it nonlinear} with a local rate of contraction given by the mean
friction $\langle \ \zeta \ \rangle $.  The time-averaged contraction rate  $\langle \ \zeta \ \rangle$
is zero on some orbits and positive on others.

Figure 2 shows two invariant tori coexisting with a limit cycle for $\epsilon =
0.42$ projected onto the $qp$-plane. The tori are produced using the initial conditions $(q,
\ p, \ \zeta) = (-2.3, \ 0, \ 0)$ and $(3.5, \ 0, \ 0)$, and their Lyapunov exponents are
$(0, \ 0, \ 0)$. The limit cycle is produced using the initial conditions $(-2.7, \ 0, \ 0)$,
and its Lyapunov exponents are $(0, \ -0.0256, \ -0.0788)$. The three objects are interlinked
as shown in Figure 2.

\begin{figure}[!ht]
\centering
\includegraphics[scale=0.4]{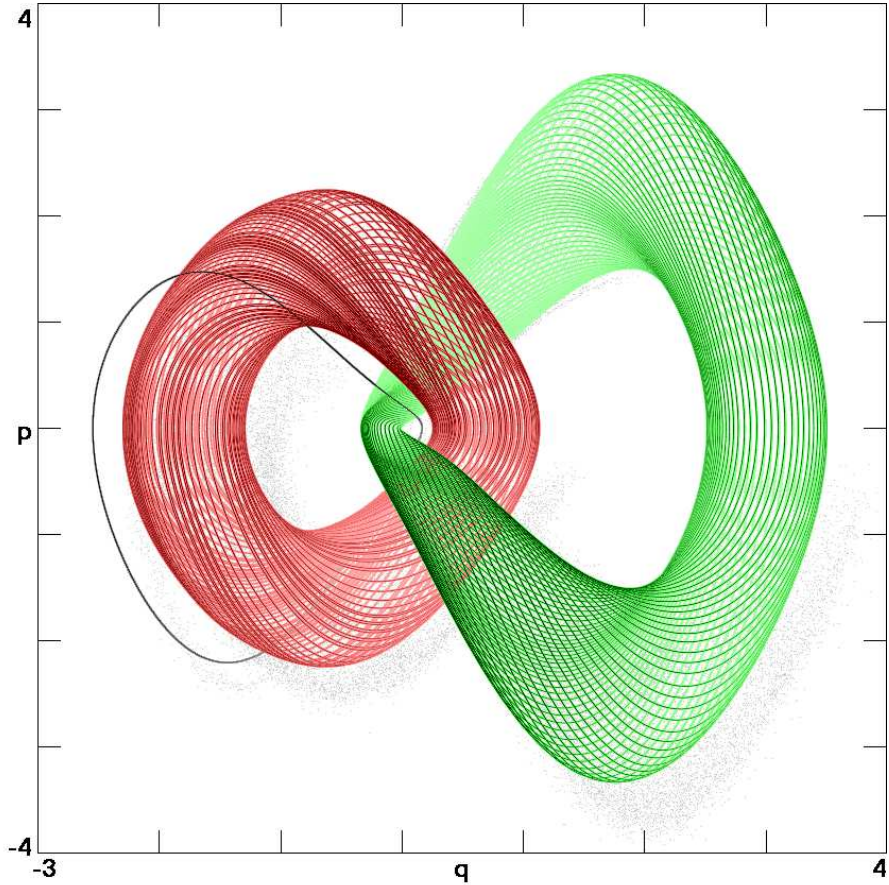}
\caption{
Three interlocked phase-space structures, two conservative ``invariant'' tori and a
 dissipative limit cycle, obtained with maximum temperature gradient $\epsilon = 0.42$, as described
in the text.  If time is reversed, with both $p$ and $\zeta$ changing signs, this structural
portrait remains unchanged.  In the $(q,p,\zeta)$ phase space the temperature is 
$1 + \epsilon \tanh(q)$ .
}
\end{figure}

These tori are only two of an infinite sequence of nested tori as shown in the $\zeta = 0$
cross section in Figure 3 where 128 initial conditions are taken uniformly over the interval
$-5 < q < 5$ with $p = \zeta = 0$. The cross section of the limit cycle is shown there as two small
red dots. The basin boundary of the limit cycle appears to coincide with the outermost torus
and extends to infinity in all directions. Orbits starting at points near the basin boundary exhibit
transient chaos before eventually converging to the limit cycle.

\begin{figure}[!ht]
\centering
\includegraphics[scale=0.6]{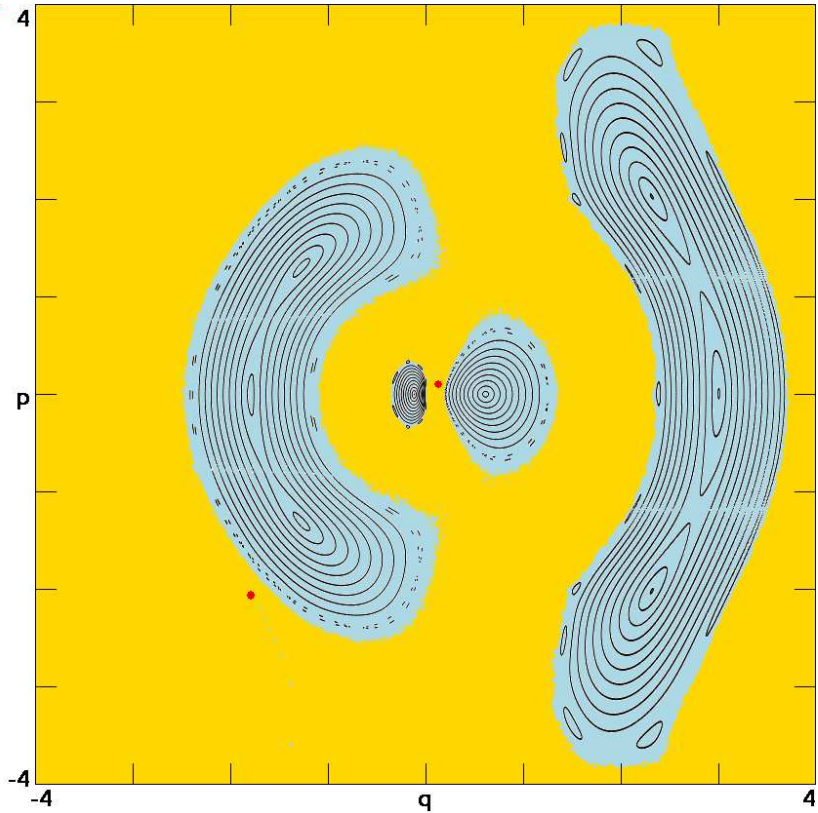}
\caption{
Detailed cross section for $\zeta = 0$ of the structure shown in Figure 2 for $\epsilon = 0.42$ with
128 initial conditions. The limit cycle is shown in cross section by the two red dots.
The dissipative region is shown in maize, the conservative in blue. 
}
\end{figure}

Because the system is trapping and invariant under the transformation $p \rightarrow -p, \ \zeta
\rightarrow -\zeta, \ t \rightarrow -t$, there is a repelling cycle symmetric with the limit
cycle toward which all orbits in the basin of the limit cycle are drawn when time is reversed,
while orbits on the tori remain on the tori.

As $\epsilon$ is decreased, the limit cycle loses its stability around $\epsilon = 0.4053_6$ and
becomes a weakly chaotic, nearly space-filling strange attractor while the tori increase in
size and complexity. At $\epsilon = 0.38$, the Lyapunov exponents in the chaotic region are
$(0.0019, \ 0, \ -0.0020)$, and the Kaplan-Yorke dimension is $2.945$. The time-averaged
dissipation is tiny as given by $\langle \ \zeta \ \rangle \simeq 1.2\times 10^{-4}$, but
decidedly nonzero. This attractor dimension is in sharp contrast to the dimensions of strange
attractors like Lorenz', and R\"ossler's.  Those three-dimensional flows have
fractal dimensions only slightly greater than $2.0$. A cross section of our $(q,\ p,\ \zeta)$
flow with $\epsilon = 0.38$ in the $\zeta = 0$ plane is shown in Figure 4. What looks like a
conservative chaotic sea is actually a weakly dissipative multifractal strange attractor with
a capacity dimension of $3.0$ and a correlation dimension of $2.09$.  The laser solutions of
Reference 10 show similar structures.

\begin{figure}[!ht]
\centering
\includegraphics[scale=0.6]{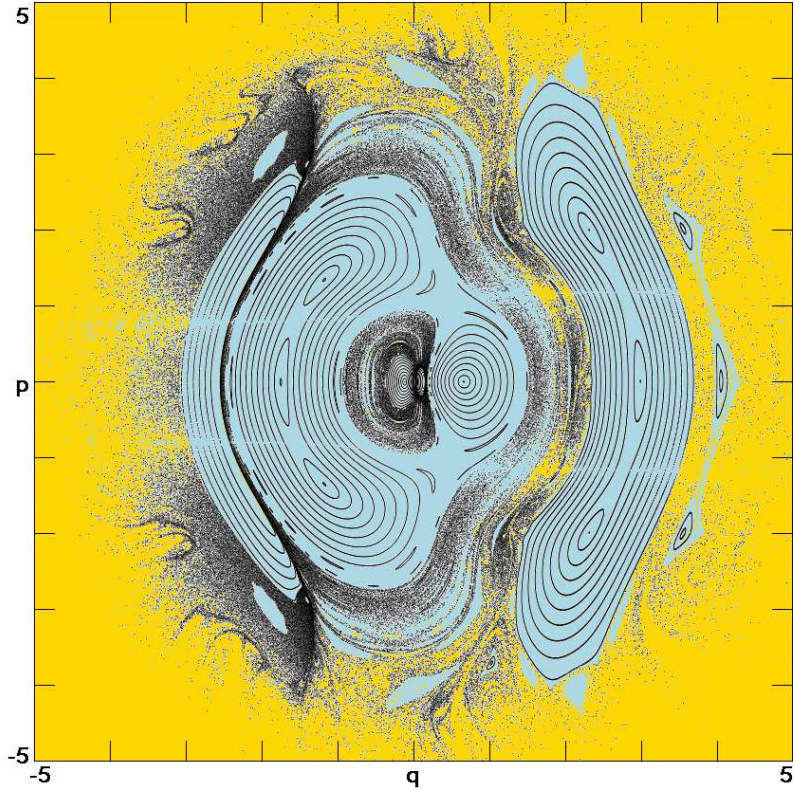}
\caption{
Cross section for $\zeta = 0$ using 128 initial conditions ( dissipative in maize and conservative
in blue ) as in Figure 3, but with a maximum temperature gradient $\epsilon = 0.38$, as described in the text.
}
\end{figure}

As $\epsilon$ is decreased further, the strange attractor becomes more chaotic (its largest
Lyapunov exponent increases) while the dissipation decreases until it vanishes at $\epsilon
= 0$ where the strange attractor becomes a chaotic sea with Lyapunov exponents of $(0.0139,\
0, \ -0.0139)$. The system is then purely conservative with invariant tori coexisting with
the chaotic sea.

The complexity of the single-thermostat oscillator system $(\ q,p,\zeta \ )$ is due to its
extreme lack of ergodicity.  This property of the model led to attempts to remedy that
lack\cite{b17,b18,b19}.  Two of these, one fixing both the second and the fourth moments
of momentum\cite{b17}, and the other fixing the oscillator second moment as well as
the second moment of the friction coefficient\cite{b19}, are applied to the oscillator problem
in the following Section.

\section{Robust Heat Transfer with a Harmonic Oscillator}

Hoover and Holian\cite{b17} used {\it two} thermostat control variables, $\zeta$ and $\xi$,
fixing both the second and the fourth long-time-averaged moments of momentum,
$\langle \ p^2,p^4 \ \rangle$ . The four equations of motions, which generate Gibbs' canonical
distribution for a fixed temperature $T$ become:
$$
\{ \ \dot q = p \ ; \ \dot p = -q - \zeta p - \xi p^3 \ ; \ \dot \zeta = p^2 - T \ ;
\ \dot \xi = p^4 - 3p^2T \ \} \ {\rm [ \ HH \ ] } \ .
$$
Both $\zeta $ and $\xi $ have Gaussian distributions\cite{b17}.  These equations of motion are
evidently ergodic, according to careful tests carried out by Posch and Hoover\cite{b6}.
Another isothermal set of four motion equations, due to Martyna, Klein, and Tuckerman\cite{b18},
are likewise thought to be ergodic:
$$
\{ \ \dot q = p \ ; \ \dot p = -q - \zeta p  \ ; \ \dot \zeta = p^2 - T - \xi \zeta \ ;
\ \dot \xi = \zeta^2 - T \ \} \ {\rm [ \  MKT \ ] } \ .
$$

To avoid confusion, but perhaps not controversy, we explain what we mean by ``ergodicity'',
which we believe to be quite similar to the Ehrenfests' concept of ``quasi-ergodicity''.  We
choose a generalized (four-dimensional) cube (or parallelepiped, or sphere, or some other
compact shape) and ask whether a trajectory started anywhere within this four-dimensional volume
will eventually come arbitrarily near any other point in the volume.  If so, ``ergodic''.  If
not, not ergodic.

Let us consider next {\it generalized} versions of the Hoover-Holian and Martyna-Klein-Tuckerman
oscillators with the hyperbolic-tangent temperature field $T = 1 +\epsilon \tanh(q) $ .
First, consider the four-dimensional, time-reversible dynamics of the doubly-thermostated,
one-dimensional HH oscillator with a maximum temperature gradient of $\epsilon = 0.40$.
Figure 5 shows those $(\ q,p \ )$ values close to equilibrium whenever $|\zeta| < 0.005$ and
$|\xi| < 0.005$. This method of obtaining a double cross section has the virtue that the density
of points in the two-dimensional plot is proportional to the density in the full
$(q,p,\zeta,\xi)$ four-dimensional phase space. Although there is considerable structure in the
plot, there is no evidence for the holes common to the three-dimensional system.  We believe
that the reason for this uniformity is the vanishing likelihood for finding a periodic orbit in
the three-dimensional space defining a three-dimensional ``Poincar\'e volume'', analogous to a
two-dimensional Poincar\'e section.  Figure 6 is the double cross section view using the MKT
chain-thermostat idea of Reference 18 with $\epsilon = 0.20$.

\begin{figure}[!ht]
\centering
\includegraphics[scale=0.4]{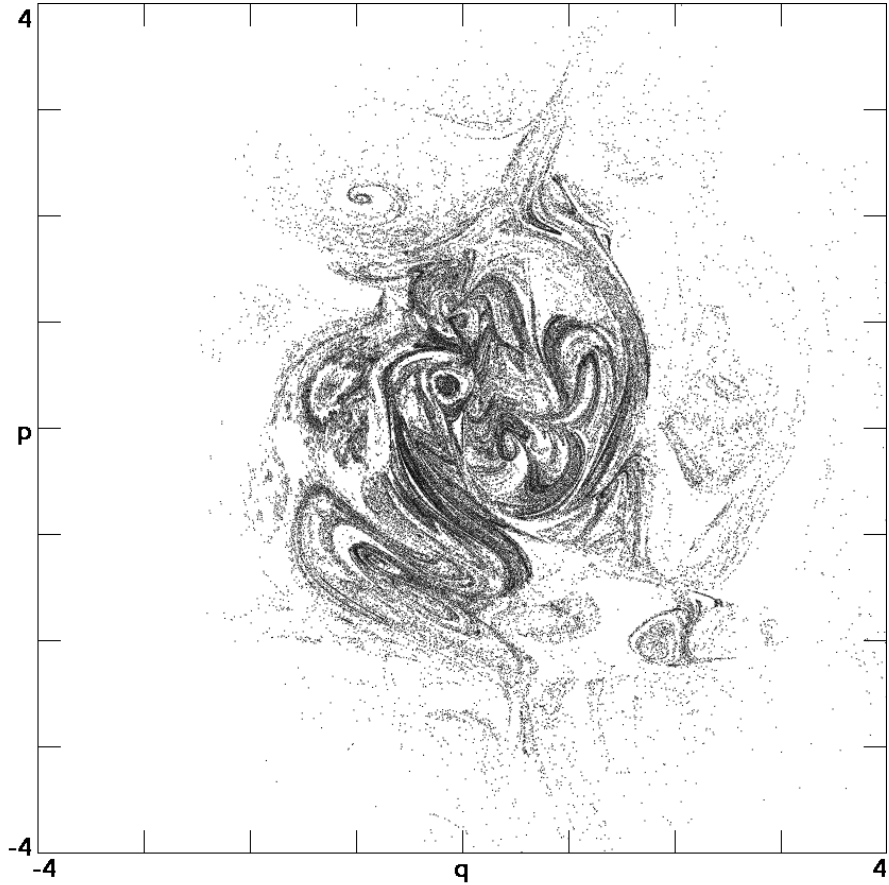}
\caption{
Double cross section ($\zeta = \xi = 0$) for the nonequilibrium version of the four HH
equations of Reference 17 with $\epsilon = 0.40$.  The three-dimensional tori are
absent here.  The Lyapunov exponents are $\{ \ 0.0878, \  0, \ -0.0084,
 \ -0.1184 \ \}$, with capacity dimension 4, Kaplan-Yorke dimension 3.687, and correlation
dimension $3.38$. 
}
\end{figure}

\begin{figure}[!ht]
\centering
\includegraphics[scale=0.4]{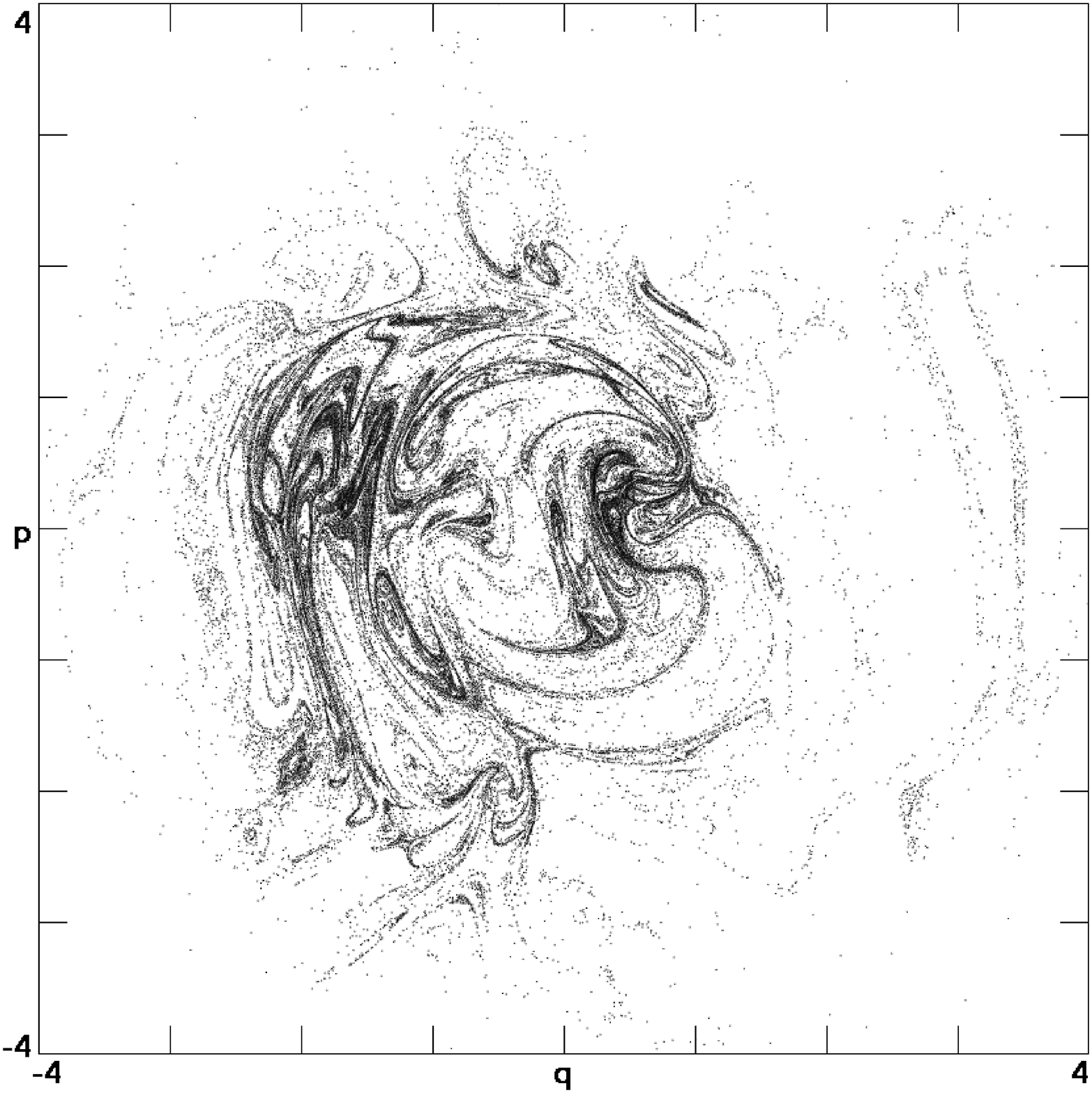}
\caption{
Double cross section ($\zeta = \xi = 0$) for the nonequilibrium version of the four MKT
equations of Reference 18 with $\epsilon = 0.20$.  The three-dimensional tori are
absent here.  The Lyapunov exponents are $\{ \ 0.0692, \  0, \ -0.0159, 
\ -0.0856 \ \}$, with capacity dimension 4, Kaplan-Yorke dimension 3.624, and correlation
dimension $3.39$.
}
\end{figure}

Because, for sufficiently small temperature gradients, the HH motion equations from Reference 17
and the MKT motion equations from Reference 18 yield space-filling ergodic solutions, the conventional
descriptions in terms of periodic orbits and saddle points are not useful. It appears that perturbing
the smoothly continuous Gaussian density,
$$
f(q,p,\zeta,\xi) \propto e^{-(q^2+p^2+\zeta^2+\xi^2)/2} \ ,
$$
away from isothermal equilibrium, by adding a further nonlinearity, leads to fractally-concentrated
ridges and depleted valleys in the density, much like the perturbations responsible for the earth's
basin and range construction. The formally unanswered question of {\it why} the four-dimensional
equations are ergodic while the three-dimensional ones are not is important, but we can only
provide an informal explanation. We encourage the mathematically-inclined reader to pursue it
with vigor and imagination. A simple explanation seems unlikely.

Very recently Patra and Bhattacharya suggested a different thermostating method, fixing {\it both}
the kinetic and configurational temperatures simultaneously by using two thermostating control
variables\cite{b19}.  We thank them for several useful emails.  Applied to the harmonic oscillator
problem, their equations of motion are:
$$
\{ \ \dot q = p - \xi q \ ; \ \dot p = -q - \zeta p \ ; \ \dot \zeta = p^2 - T \ ; \
\dot \xi = q^2 - T \ ; \ T \equiv 1 + \epsilon \tanh (q) \ \} \ .
$$
From the standpoint of control theory the Patra-Bhattacharya thermostating idea
is not far from ( a rewritten form of ) the laser equations (2.3) of Politi, Oppo, and Badii :
$$
\{ \ \dot q = p - \zeta q \ ; \ \dot p = -q - \zeta p \ ; \ \dot \zeta = q^2 + p^2 - T \ \} \ . 
$$
It is interesting to see that the Patra-Bhattacharya set of four equations shows all the complexity
of the Nos\'e-Hoover equations but lacks the space-filling ergodicity of the two other four-dimensional
systems investigated here. With $\epsilon = 0$ the equilibrium Patra-Bhattacharya equations show
a strong correlation between the oscillator coordinate and momentum, with $\langle \ q^2p^2 \ \rangle
\simeq 1.41$ rather than unity.  Numerical work with these equations is complicated by the fixed
points at $\epsilon = 0$, $(q,p) = (\pm 1,\pm 1)$ , which can be removed by using other moments.

Sprott\cite{b20} has discovered another time-reversible three-dimensional dynamical system which
displays coexisting conservative and dissipative regions:
$$
\{ \ \dot x = y + 2xy + xz \ ; \ \dot y  = 1 - 2x^2 + yz \ ; \ \dot z = x - x^2 - y^2 \ \} \ .
$$ 
There are now a wealth of such time-reversible dynamical systems which exhibit thermodynamic
characteristics.

\section{Conclusions and Recommendations}

A re\"investigation of the decades-old Nos\'e-Hoover-Posch-Sprott-Vesely work is well warranted
by the more-recent advances in processor speeds.  The calculations described here
can be carried out on a laptop computer in a clock time measured in a few minutes or hours.
In 1984 such problems required multi-million-dollar Cray Supercomputers.

The simplest of the Nos\'e-Hoover nonequilibrium simulations show a qualitative difference between
tori and limit cycles.  Only the limit cycles can support dissipation.  This dissipation is shared
with the space-filling strange attractor that surrounds the various quasi-periodic solutions
of the motion equations.  How do the tori resist dissipation?  A clear explanation is still lacking.

The difference between the complexity of the three-dimensional systems and the relative simplicity of
two of the four-dimensional ones (but not the Patra-Bhattacharya equations) is striking. This observation
suggests that quasi-periodic structures are {\it relatively rare} in four-equation two-or-three-dimensional
Poincar\'e cross sections.  On the other hand quasi-periodic structures are {\it commonplace} in three-equation
two-dimensional Poincar\'e cross sections.  Quasi-periodic solutions are {\it inevitable} in two dimensions.
The question remains as to {\it why} the Patra-Bhattacharya oscillator shows quasi-periodic toroidal behavior
for even the smallest values of $\epsilon$.

These results should provide grist for the mathematicians' mills for some time.  We also recommend them to
students for further rewarding study.

\newpage

\section{Acknowledgment}
We thank Antonio Politi, Karl Travis, and the referees for their suggestions and careful readings of
the manuscript.  One referee asked that we include the clarifying explanation that ``homoclinic cycles partition
the phase space into its conservative and dissipative parts, though the relevance of such cycles to
macroscopic thermodynamics is still unexplored''.

\end{document}